\documentclass[twocolumn, tighten, times]{aastex61}
\usepackage{graphicx}
\usepackage{epstopdf}

\usepackage{color}
\usepackage{soul}

\newcommand{\um}{$\mu m$}
\renewcommand\baselinestretch{1.3}
\shorttitle{Mass dependence of structure and SFR for galaxies at $0.5<z<2.5$}
\shortauthors{Gu et al.}

\begin{document}

\title{The mass dependence of structure, star formation rate, and mass assembly mode \\ at $0.5<z< 2.5$}

\correspondingauthor{Qirong Yuan}
\email{yuanqirong@njnu.edu.cn}

\author{Yizhou Gu}
\affil{School of Physics Science and Technology, Nanjing Normal University,
Nanjing  210023, China;  
yuanqirong@njnu.edu.cn}

\author{Guanwen Fang}
\altaffiliation{Guanwen Fang and Yizhou Gu contributed equally to this work}
\affil{Institute for Astronomy and History of Science and Technology, Dali University, Dali 671003, China; wen@mail.ustc.edu.cn
} 

\author{Qirong Yuan}
\affil{School of Physics Science and Technology, Nanjing Normal University,
Nanjing  210023, China;  
yuanqirong@njnu.edu.cn}

\author{Shiying Lu}
\affil{School of Physics Science and Technology, Nanjing Normal University,
Nanjing  210023, China;  
yuanqirong@njnu.edu.cn}

\author{Feng Li}
\affil{School of Physics Science and Technology, Nanjing Normal University,
Nanjing  210023, China;  
yuanqirong@njnu.edu.cn}

\author{Zhen-Yi Cai}
\affil{CAS Key Laboratory for Research in Galaxies and Cosmology, Department of Astronomy, University of Science and Technology of China, Hefei, Anhui 230026, China}
\affil{School of Astronomy and Space Science, University of Science and Technology of China, Hefei 230026, China}

\author{Xu Kong}
\affil{CAS Key Laboratory for Research in Galaxies and Cosmology, Department of Astronomy, University of Science and Technology of China, Hefei, Anhui 230026, China}
\affil{School of Astronomy and Space Science, University of Science and Technology of China, Hefei 230026, China}

\author{Tao Wang}
\affil{Institute of Astronomy, University of Tokyo, 2-21-1 Osawa, Mitaka, Tokyo 181-0015, Japan}



\begin{abstract}
To investigate the mass dependence of structural transformation and star formation quenching, we construct three galaxy samples using massive ($M_* > 10^{10} M_{\sun}$) red, green, and blue galaxy populations at $0.5<z<2.5$ in five 3D--{\it HST}/CANDELS fields.  
The structural parameters, including effective radius ($r_{\rm e}$), galaxy compactness ($\Sigma_{1.5}$), and  second order moment of 20\% brightest pixels ($M_{20}$) are found to be correlated with stellar mass. S\'{e}rsic index ($n$), concentration ($C$), and Gini coefficient ($G$) seem to be insensitive to stellar mass. 
The morphological distinction between blue and red galaxies is found at a fixed mass bin, suggesting that quenching processes should be accompanied with transformations of galaxy structure and morphology. 
Except for $r_e$ and $\Sigma_{1.5}$ at high mass end,  structural parameters of green galaxies are intermediate between red and blue galaxies in each stellar mass bin at $z < 2$, indicating green galaxies are at a transitional phase when blue galaxies are being quenched into quiescent statuses. The similar sizes and compactness for the blue and green galaxies at high-mass end implies that these galaxies will not appear to be significantly shrunk until they are completely quenched into red QGs. 
For the green galaxies at $0.5<z<1.5$, a morphological transformation sequence of bulge buildup can be seen as they are gradually shut down their star formation activities, while a faster morphological transformation is verified for the green galaxies at $1.5<z<2.5$. 
\end{abstract}

\keywords{galaxies: evolution - galaxies: high-redshift - galaxies: structure }


                 
\section{Introduction} \label{sec:intro}

It is well known that the dichotomy of galaxy populations, which has been revealed in the local universe (e.g, \citealt{Strateva+01, Baldry+04}), exists up to $z\sim 2.5$ (e.g, \citealt{Brammer+09, Xue+10, Whitaker+11}). 
In the color$-$magnitude diagram, abundant quiescent galaxies and a small amount of dusty star-forming galaxies are distributed  along with a linear sequence, which is called ``red sequence'' (RS). 
Conversely, star-forming galaxies (SFGs) mainly occupy an extended region, which is called ``blue cloud'' (BC). As star formation activities in SFGs are gradually ceased, the galaxies would move across an intermediate zone between BC and RS,  so-called ``green valley'' (GV). 
From the view angle of morphologies and structural parameters, such as S\'{e}rsic index ($n$), effective radius ($r_{\rm e}$), compactness ($\Sigma_{1.5}$), Gini coefficient ($G$), concentration ($M_{20}$), and the second order moment of the 20\% brightest pixels, the GV galaxies are proved to be at the transitional phase when SFGs are being quenched into red quiescent galaxies (QGs) (\citealt{Pandya+17,Gu+18}). 

The evolution of massive galaxies is, in a sense, a process of mass assembly. Stellar mass is a crucial physical property of a galaxy that is naturally thought to be related to other physical properties. 
It is well known that for SFGs the star formation rate (SFR) is found to be tightly correlated with the stellar mass, which is called the ``star forming main sequence'' (SFMS). 
According to the SFR--M$_*$ relation, SFGs are thought to grow steadily and smoothly, reaching a balance between gas inflow and new star formation (e.g., \citealt{Elbaz+07, Dekel+13, Mancuso+16}). 
The SFMS  can be sensed by its presence to higher redshifts, even up to $z \sim 4$ (e.g., \citealt{Elbaz+07,Salim+07, Wuyts+11a,Speagle+14,Whitaker+14,Tomczak+16}).
Moreover, it has also been observed that galaxy properties change significantly once they deviate from the SFMS on the SFR--M$_*$ plane (\citealt{Wuyts+11b, Brennan+17, Lee+18}). Considering the strong correlation between galaxy structural properties and deviation from the SFMS, it is highly possible that the stellar mass assembly has simultaneously affected their structures and star formation properties.

Besides the SFR--M$_*$ relation, stellar mass is thought to be related to other physical properties. \citet{Kauffmann+03} found strong correlations between star formation history, stellar mass, and structural parameters on the basis of spectral analysis using a large sample of SDSS galaxies. 
The massive SDSS galaxies with $\rm M_* > 3 \times 10^{10} M_{\sun}$ are found to be more concentrated and dominated by older stellar populations, whereas the low-mass galaxies are likely to have lower concentrations and younger stellar ages. Many previous studies discuss the mass dependence of galaxy structure between QGs and SFGs (e.g., \citealt{vdW+14, Lang+14, Whitaker+15, Whitaker+17, ArgudoF+18, Mowla+18, Matharu+19, Miller+19}). 
For two main classes of galaxies at high redshifts,  their stellar masses have also been found to be correlated with  structure parameters  (e.g., the size -- mass relation and the density -- mass relation) (\citealt{Barro+17,Lee+18}). 
The QGs and SFGs show different dependencies of size on the stellar mass up to $z \sim 3$ (\citealt{Shen+03, Daddi+05, Trujillo+07, Newman+12, Barro+13, vdW+14}).  
Compared with the SDSS galaxies in the local universe, the high-$z$ galaxies  in the early epoch tend to be smaller and more compact.
 \citet{Barro+17} unveiled that the surface densities within the core region ($R<1$kpc) ($\Sigma_1$) and at effective radius ($\Sigma_e$) show different tight correlations with the stellar mass, which manifests a synchronous mass growth of disk and core for SFGs. Moreover, a ``downsizing'' picture is expected that more massive galaxies complete the bulk of star formation earlier than galaxies with lower stellar masses (\citealt{Cowie+96, Goncalves+12}). In a word, stellar mass is an important quantity to determine the structure and star formation history for a galaxy. 

Recently, \citet{Gu+18} have constructed a large sample of massive red, green, and blue galaxies with $M_\sun \geqslant 10^{10} M_*$ at $0.5\leqslant z \leqslant 2.5$ in five fields of 3D--{\it HST}/CANDELS. By adopting our redshift dependent definitions of RS, GV, and BC galaxies via the extinction-corrected rest-frame (U$-$V) color versus the stellar mass diagram, we carried out an investigation on morphologies, AGN fractions, dust content, and environments for the red, green, and blue galaxy populations at $0.5<z<2.5$ in 3D--{\it HST}/CANDELS fields (\citealt{Grogin+11, Koekemoer+11,Skelton+14}), and unveiled  a ``downsizing'' quenching picture (\citealt{Gu+18}).  

In this work, we set out to provide empirical relations between stellar mass and other galaxy properties for the three subsamples of BC, GV, and RS galaxies at various redshift bins. 
Based on our large sample, we should figure out how the structures and star formation rates depend on the stellar mass for three galaxy populations, which will shed light on how galaxies with various stellar masses alter their structures during the cessation of their star formation activities at various redshifts.
We should also present a morphological map in the SFR--M$_*$ plane, and estimate the time scale of quenching over cosmic time. 
To avoid the potential bias induced by the mass selection, three mass-matched samples of the red, green, and blue galaxies will be constructed to verify the scenario that the green galaxies are at a transitional phase when the blue galaxies are being quenched into the quiescent ones.

The structure of our paper is organized as follows. The 3D--{\it HST}/CANDELS data set and our sample construction are described in Section \ref{sec:data}. 
We examine the mass dependence of the morphological parameters in Section \ref{sec:mass-dependence}.  
Galactic morphology and the link to its location in the SFR--M$_*$ plane are presented in Section \ref{sec:SFR-M-trans}.  
In Section \ref{sec:discuss}, structure parameters varying with mass and redshift, the average transition time scale, and the mass-matched samples are discussed. 
Finally, a summary is given in Section \ref{sec:summary}. 
Throughout our paper, we adopt the cosmological parameters as following: $H_0=70\,{\rm km~s}^{-1}\,{\rm Mpc}^{-1}$, $\rm \Omega_m=0.30$, and $\Omega_{\Lambda}=0.70$.

\section{Data} \label{sec:data}
\subsection{CANDELS and 3D--{\it HST}}

Our work is based on the data from CANDELS (\citealt{Grogin+11, Koekemoer+11}) and 3D--{\it HST} (\citealt{Skelton+14}) programs, which provide abundant multi-wavelength photometric data at wavelengths of 0.3$-$8.0 $\mu m$ from many space- and ground-based telescopes in five different fields: AEGIS, COSMOS, GOODS-N, GOODS-S, and UDS. The total sky coverage of these five fields is over 900 arcmin$^2$, which is helpful to minimize the effect of cosmic variance.

Photometric redshifts, rest-frame colors, and stellar population parameters have been derived for the 3D--{\it HST}/CANDELS (\citealt{Skelton+14}) by fitting the spectral energy distribution (SED) of each galaxy with  a linear combination of seven templates via the EAZY code (\citealt{Brammer+08}). 
In this work, we prefer to take the spectroscopic redshift if available. For the remaining galaxies without spectroscopic redshift, we adopt the photometric redshifts derived from \cite{Skelton+14}. 
To better estimate the dust attenuation ($A_V$) for the SFGs at high redshifts and following \cite{WangT+17}, we construct galaxy templates by taking the \cite{Ma+05} stellar population synthesis (SPS) models which add the contribution of asymptotic giant branch stars, instead of taking the \cite{BC03} SPS.
Assuming an exponentially declining star forming history with e-folding timescales ranging from $10^8$ to $10^{10}$ yr, the \cite{Calzetti+00} extinction law, and the \cite{Kroupa+01} initial mass function (IMF), we re-estimate the stellar population parameters using the FAST code (\citealt{Kriek+09}).  The derived values of the stellar mass and the dust attenuation will be adopted to construct the samples of red, green, and blue galaxies in Section \ref{sec:sample}.

\begin{figure*}
\centering
\includegraphics[scale=0.90]{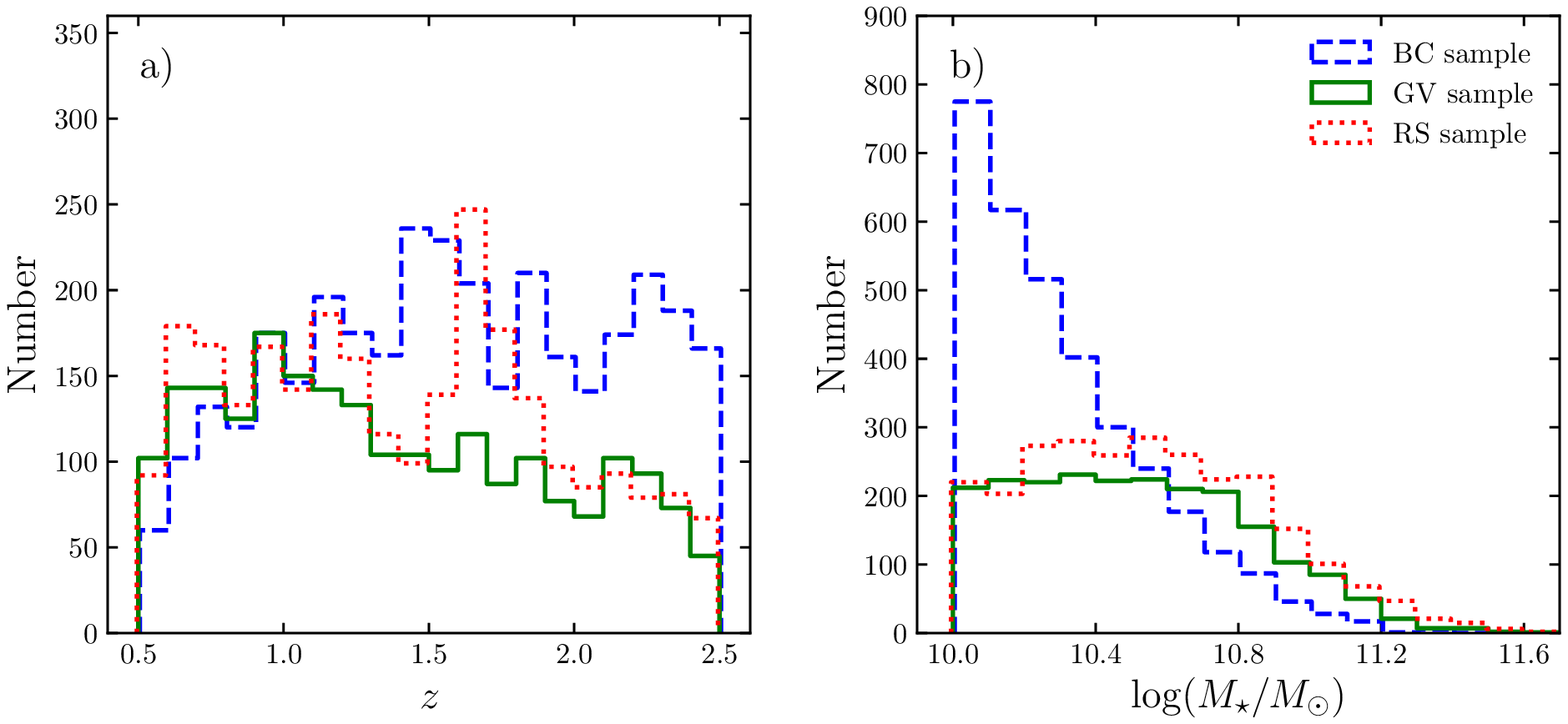}
\caption{The redshift (left panel) and the stellar mass (right panel) distributions for the RS, GV, and BC galaxies at $0.5<z<2.5$, denoted by the red dotted, the green solid, and the blue dashed lines, respectively. 
\label{fig01}}
\end{figure*}

\begin{figure}
\centering
\includegraphics[scale=0.90]{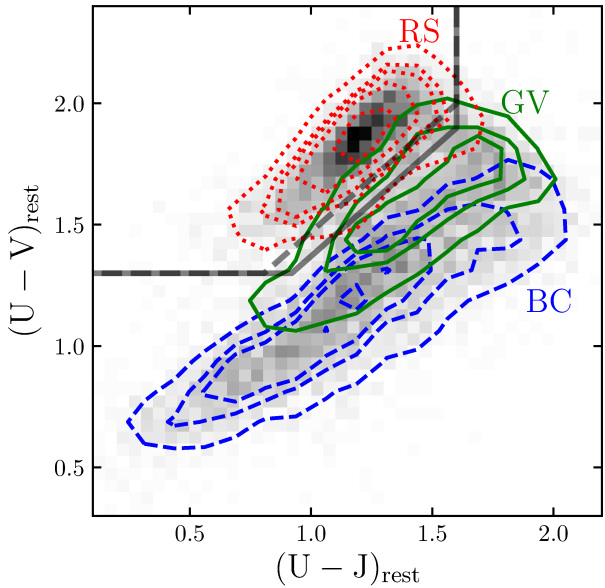}
\caption{The rest-frame color-color diagram for massive galaxies at $0.5<z<2.5$, where the red dotted, the green solid, and the blue dashed contours represent the RS, GV, and BC galaxy populations, respectively. 
The widely adopted \cite{Williams+09} criteria separating SFGs from QGs at $0.5<z<1.0$ and $z>1.0$ are shown as the gray dashed and solid lines, respectively. 
\label{fig02}}
\end{figure}

\subsection{Structural Parameters and Morphological Classification}
Morphologies and structures of galaxies are packed with the information about the kinematics and evolutionary history (e.g., \citealt{KK+04}). 
For a more comprehensive analysis of morphologies and structures, it is helpful to combine the S\'{e}rsic parameters, the nonparametric measurements, and the deep learning classifications together.

\subsubsection{Parametric Measurements}
The surface brightness of galaxies is widely parameterized by the single S\'{e}rsic profile with an effective radius ($r_{\rm e}$) and a S\'{e}rsic index ($n$).
The star-forming galaxies tend to be disk-like with $n\sim 1$, whereas the quiescent galaxies are more cuspidal (e.g., \citealt{Wuyts+11b, Bruce+12, Bell+12, Conselice+14}). 
The S\'{e}rsic index ($n$), the effective radius ($r_{\rm e}$), and the axis ratio ($b/a$) are measured using the {\it HST}/WFC3 $J_{\rm F125W}$ and $H_{\rm F160W}$ photometries.  
As presented in \cite{vdW+12}, these structural parameters are well measured with GALFIT (\citealt{PengCY+02}). 
In this work, the rest-frame optical morphologies are traced by J-band images at $0.5<z<1.5$ and by H-band images at $1.5<z<2.5$. 
It is found that accurate measurements --- to 10\% or better --- of all structural parameters can typically be obtained for galaxies with $H_{\rm F160W} < 23$, with comparable fidelity for size and shape measurements to $H_{\rm F160W} \sim 24.5$ \citep{vdW+12}. 

\subsubsection{Nonparametric Measurements}
The model-independent nonparametric measurements are advantageous for galaxies at high redshifts, where a larger fraction of them may be irregular and do not have distinct centers (see the review by \citealt{Conselice+14}). 
The concentration index ($C$) describes the aggregation degree of the surface brightness distribution of a galaxy (\citealt{Abraham+94}).
The Gini coefficient ($G$) is a statistical coefficient to quantify the uniformity of light distribution (\citealt{Lotz+04}).
The second-order moment of the 20\% brightest pixels ($M_{20}$) traces the substructures in a galaxy, such as bars, spiral arms, and multiple cores (\citealt{Lotz+04}). 

We have performed measurements of these nonparametric parameters for all galaxies in the five 3D-{\it HST}/CANDELS fields, using the Morpheus software developed by \cite{Abraham+07}. Our measurements have been used and tested by many previous works \citep{Kong+09, Wang+12, Fang+15, Gu+18}. 
It is shown that the nonparametric measurements depend strongly on the Signal-to-Noise ratio per pixel (S/N), especially for S/N $\leq 2$ \citep{Lotz+04, Lisker+08}. 
One of our selection criteria, $\tt use\_phot = 1$, ensures a reliable detection in $H_{\rm F160W}$ with S/N $> 3$ (see Section \ref{sec:sample}). 
In addition, it is found that there is good correspondence between the visual morphological types and their distributions in the $G -M_{20}$ diagram. 
Therefore the nonparametric measurements for our sample are reliable and will not suffer from the S/N effect.

\subsubsection{Morphological Classifications By Deep Learning}
The aforementioned structural parameters quantifying galactic structures reduce the complexity of galaxy classification. 
However, it might neglect an enormous amount of information contained in pixels. 
The deep-learning algorithm can use all of the pixels as the parameter space to quantify the morphological type. 
Based on the $H$-band images for the five CANDELS fields, a catalog of morphological classification based on deep-learning algorithm has been given by \cite{HCPG+15}, which will be adopted for further analysis. It includes the probabilities of having spheroid ($f_{\rm spheroid}$),  having disk ($f_{\rm disk}$), having some irregularities ($f_{\rm irr}$), being a point source ($f_{\rm PS}$), or being unclassifiable ($f_{\rm Unc}$). 
With the selection criteria in \cite{HCPG+15}, the massive galaxies can be classified into four typical morphological classes: (1) spheroidal galaxies (bulge dominated), (2) early-type disk galaxies (bulge dominated and having a disk), (3) late-type disk galaxies (disk dominated), and (4) irregular galaxies (including the irregulars  and mergers). We specify these four classes as SPH, ETD, LTD, and IRR, for short. 
By cross-matching the morphological catalog of \cite{HCPG+15}, about 90\% of the massive galaxies in our sample have been classified into these four typical morphologies. 
We further classify the remaining galaxies with eyeballing inspection. 

\subsection{Star Formation Rates}
Considering contributions from both UV and IR emissions, the SFRs of galaxies in the five CANDELS fields have been estimated by \cite{Whitaker+14}, assuming that the IR emission of galaxies ($\rm L_{IR}$) originates from the dust heated by the obscured UV light emitted by young, massive stars. 
By default, there is an estimate of $\rm SFR_{UV}$ for each galaxy derived from its SED-based rest-frame NUV luminosity at 2800\AA.
According to a ladder of SFR indicator based on the prescription given by \cite{Wuyts+11a}, the SFR contributed by heated dust can be derived using a single MIPS measurement at 24 $\mu $m.  
Thus, by adding its IR contribution to that of the obscured UV luminosity ($\rm L_{UV}$), the total SFR of galaxies can be calculated.  
Using the conversion of \cite{Bell+05} and scaling to the \cite{Chabrier+03} IMF, the SFRs can be derived by
\begin{eqnarray}
{\rm SFR_{UV+IR}}[M_\sun~{\rm yr}^{-1}]=1.09\times10^{-10}(L_{\rm IR}+2.2L_{\rm UV})/L_{\sun},
\end{eqnarray}
where $L_{\rm IR}$ is the integrated luminosity at 8-1000 \um ,  and $L_{\rm UV}$ represents the luminosity from 1216\AA\ to 3000\AA\ in the rest frame. $L_{\rm UV}$ can be estimated by the rest-frame continuum luminosity at 2800\AA : $L_{\rm UV}=1.5 L_{2800}$, where the factor of 1.5 accounts for the UV spectral shape of a 100 Myr old population with a constant SFR. 

If the MIPS 24 $\mu m$ data are unavailable, the effect of dust attenuation on the UV-based SFR can be corrected by assuming the \cite{Calzetti+00} dust attenuation curve: 
\begin{equation}
 {\rm SFR_{UV, corr}}[M_\sun~{\rm yr}^{-1}]={\rm SFR_{UV}}\times10^{0.4\times 1.8 \times A_V},
\end{equation}
where ${\rm SFR_{UV}}=3.6 \times 10^{-10} \times  L_{2800}/L_{\sun}$ assuming a \cite{Chabrier+03} IMF (\citealt{Wuyts+11a}), $A_V$ is the SED-based optical attenuation yielded by the FAST code, and the factor of 1.8 converts $A_V$ to that at 2800\AA\ when adopting the \cite{Calzetti+00} attenuation curve. 

\subsection{Sample Selection}\label{sec:sample}
We construct a sample of massive galaxies with reliable detection ($\tt use\_phot = 1$) and $\log (M_*/M_\sun) \geqslant 10$ in the five 3D--{\it HST}/CANDELS fields (\citealt{Gu+18}). The flag of $\tt use\_phot = 1$ assigned to a galaxy indicates
that the source (1) is not a star and not close to a bright star; (2) is well exposed, namely, requiring that each object securely detected is covered
by at least two individual exposures in each of the F125W and F160W bands; (3) has a signal-to-noise ratio $S/N >3$ in the F160W image; and (4) has a passable photometric redshift fit and a “non-catastrophic” stellar population fit (See Section 3.8 in \citealt{Skelton+14}). Our sample contains 8244 massive galaxies at $0.5<z<2.5$, and the majority ($\sim 99\%$) of galaxies have $H_{\rm F160W} < 23.5$. Since we only focus on the massive galaxies with reliable detection, most galaxies in our sample are bright enough ($H_{F160W} < 24.5$), ensuring reliable measurements of the structural parameters.
The fainter galaxies with $H_{F160W} > 24.5$ only make up a very small portion ($\sim 1\%$), and we get rid of these galaxies from our sample. 

After taking dust attenuation into account, \cite{WangT+17} found that the intrinsic rest-frame colors of SFGs and QGs depend on the stellar mass and the redshift.  The following color-based separation criteria are built to define the BC, GV, and RS galaxy populations:
\begin{eqnarray} \nonumber
 (U - V )_{\rm rest} - \Delta A_V = 0.126 \log (M_*/M_{\sun}) + 0.58 - 0.286z; \,\, \nonumber  \\
\nonumber
 (U - V )_{\rm rest} - \Delta A_V = 0.126 \log (M_*/M_{\sun})  - 0.24 - 0.136z, \,\, \nonumber
\end{eqnarray}
where $\Delta A_V=0.47 A_V$ is the extinction correction of rest-frame $U - V$ color, and the correction factor of 0.47 is calculated according to the \cite{Calzetti+00} extinction law. 
Adopting the above selection criteria, our massive galaxies at $0.5 \leqslant z\leqslant 2.5$ are divided into blue, green, and red galaxies, corresponding to the three galaxy populations in the BC, GV, and RS, respectively.  

At first, we exhibit the redshift (left panel) and the stellar mass (right panel) distributions for the blue, green, and red galaxies in Figure \ref{fig01}. 
As shown, the blue galaxies are predominate at high redshifts ($z>1.5$), and the mass distribution of blue galaxies is significantly different from those of green and red galaxies. 
A larger proportion of blue galaxies can be found at the lower mass end. 
Since the morphology and structure of galaxies may depend on the stellar mass, 
the potential mass effects on the statistics of galaxy morphology and structural parameters might be induced due to the different mass distributions of the three galaxy populations, which has been mentioned by \cite{Pandya+17}. 

The {\it UVJ} color diagram is widely used to separate galaxies into star-forming and quiescent (e.g., \citealt{Williams+09, Straatman+16, Fang+18}). 
Figure \ref{fig02} shows the rest-frame $U-V$ versus $V-J$ colors for the BC, GV, and RS populations. The widely adopted cut-off rules (\citealt{Williams+09}) are illustrated as the dashed and solid lines for $0.5<z<1.0$ and $z>1.0$, respectively. We find that the wedge-shaped quiescent region is dominated by the red galaxies, whereas the rest star-forming region is mainly occupied by the blue SFGs. 
Furthermore, the green galaxies are scattered around the separation lines, just between the crowded regions of the blue and red galaxies. 
In general, our separation criteria based on the dust-corrected colors work very well in population selection. The dust-corrected $U-V$ colors may trace the intrinsic colors of galaxies.

\begin{figure}
\centering
\includegraphics[scale=0.90]{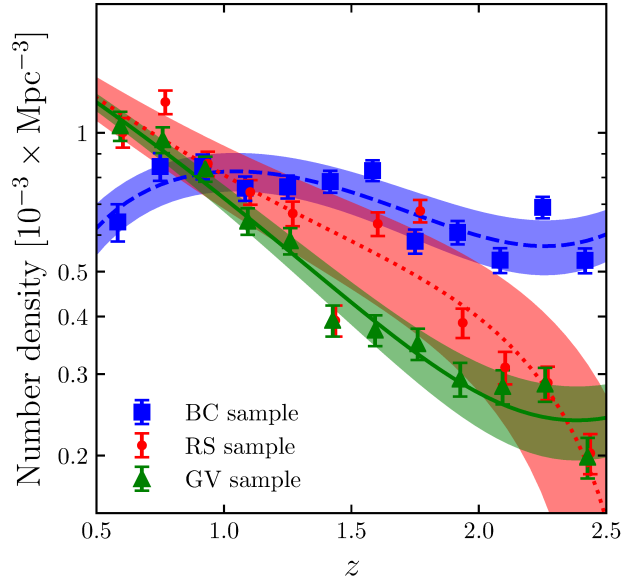}
\caption{The number densities of the BC (blue), GV (green), and RS (red) galaxies as a function of redshift, superimposed by the corresponding cubic polynomial fittings (lines) and uncertainties (shadows). 
\label{fig03}}
\end{figure}

Figure \ref{fig03} shows the comoving number densities of the massive blue, green, and red galaxies as a function of redshift, where we apply a minimum systematic fractional uncertainty of 5\%, following \cite{Muzzin+13} and \cite{Pandya+17}. 
We also apply the cubic polynomial fittings for the three galaxy populations, which would be used to estimate
the transition time-scale on average in Section \ref{sec:timescales}.
In general, the number densities of the green and red galaxies increase with cosmic time. 
The number density of blue galaxies increases at $z > 1$ and shows a deceasing trend since $z \sim 1$. 
A similar trend has been shown by \cite{Pandya+17}, who classified galaxies into SFGs, transition galaxies, and QGs based on the specific star formation rate defined as ${\rm SSFR \equiv SFR}/M_*$.

\section{Mass dependence of galactic structure}\label{sec:mass-dependence}
In this section, we focus on the correlations between the stellar mass and various structural parameters, including parametric measurements (i.e., $r_{\rm e}$ and $n$), galaxy compactness ($\Sigma_{1.5}$), and non-parametric ones (i.e., $C$, $G$, and $M_{20}$), for the BC, GV and RS galaxies. The structural parameters versus mass diagrams for the three galaxy populations in four redshift bins are shown in Figure \ref{fig04}.

To explore the mass dependence of galactic structural properties and its cosmic evolution, our red, green, and blue galaxies are further divided into three stellar mass bins with an interval of 0.4 dex (i.e., $10.0 \leqslant\log(M_*/M_\sun)< 10.4$, $10.4 \leqslant\log(M_*/M_\sun) < 10.8$, and $\log(M_*/M_\sun) \geqslant 10.8$), and into four redshift bins with an interval of 0.5 (i.e., $0.5\leqslant z < 1$, $1\leqslant z < 1.5$, $1,5\leqslant z < 2$, and $2\leqslant z\leqslant 2.5$). 
For each specific structural parameter, we would use the median (symbol) and the 25th to 75th percentiles (error bar) of each galaxy population to discuss its correlation with the stellar mass and evolution with the redshift.
To show how the structural parameters of red, green, and blue galaxies vary with the stellar mass and the redshift, we apply linear fittings to the structural parameters as a function of the stellar mass. It is noted that the fitting results for the blue galaxies are mainly determined by galaxies with $10^{10}<M_*/M_\sun<10^{10.8} $, since a very small fraction ($\sim$ 5\%) of the blue galaxies are found with $M_*>10^{10.8} M_\sun$ (see the right panel of Figure \ref{fig01}). And the fitting results for the blue galaxies at the high-mass end are largely driven by the extrapolation of trends seen at lower mass.


Inspired by \cite{Lee+18}, we also introduce the Spearman's rank correlation coefficient, $r_s$, to describe the monotonic relationship between the morphological parameter and the stellar mass. 
 The coefficients for various subsamples are given at the right bottom of each panel in different colors. Throughout the paper, we apply the following standards: 
 $|r_s| \leqslant 0.2$ means no significant correlation;  
 $0.2<|r_s| \leqslant 0.4$ means a weak correlation;  
 $0.4<|r_s| \leqslant 0.6$ means a clear or moderate correlation; 
 $0.6<|r_s| \leqslant 0.8$ indicates a strong correlation; and 
 $0.8<|r_s|\leqslant 1.0$ indicates an extremely strong correlation.


\subsection{The Size--Mass Relation}
For a given redshift and mass bin, the star-forming galaxies are significantly larger in size than quiescent systems in the rest-frame optical band (e.g., \citealt{Shen+03, Newman+12, Lang+14, vdW+14, Allen+17}). In other words, it is found that the star-forming and quiescent galaxies follow significantly different size--mass relations. 

The size--mass distributions for the three galaxy populations within four redshift regions are shown in the first line of Figure \ref{fig04}. 
On average, the effective radius tends to increase with the stellar mass and cosmic time for the three galaxy populations. The size distinction between blue and red galaxies is significant in each redshift bin; the distinction tends to be smaller with increasing the stellar mass. At high stellar masses ($M_*> 10^{10.8}M_\sun$), the size distinction among the three galaxy populations tends to become subtle at all redshifts. 

The blue galaxies show a weak size--mass correlation which is indicated by the smallest slope ($\sim 0.2$) and correlation coefficient ($r_s = 0.20-0.26$) over the whole redshift range. This supports the scenario in which the blue galaxies assemble stellar mass and increase their radii progressively \citep{vD+15}. 
Generally, the green galaxies have intermediate slopes compared to the other two populations. The size difference between blue and red galaxies tends to be more remarkable at lower masses and redshifts.  
The green galaxies with $ M_*<10^{10.4} M_\sun$ have smaller sizes than the blue galaxies at $0.5<z<2.0$, and they obey almost the same size-mass relation at $z > 2.0$. 
Compared with the green and blue galaxies, the red galaxies show the steepest slope and the highest correlation coefficient ($r_s=0.32-0.66$) for each redshift bin.




\subsection{The S\'{e}rsic Index–-Mass Relation}{\label{sec:size-mass}}
The S\'{e}rsic index versus mass diagrams for the three galaxy populations in four redshift bins are shown in the second line of Figure \ref{fig04}. In general, the structural difference between blue galaxies and red galaxies are significant. The light profile of quiescent galaxies is cuspier (higher $n$) than star-forming systems over the whole mass range. 
At $0.5<z< 1.5$, the three galaxy populations are found to have an increasing trend of the S\'{e}rsic index along the stellar mass, indicating that mass assembly leads to a clear buildup of bulge. This tendency has been found by other works (e.g., \citealt{Lang+14}). 
According to the correlation coefficients ($r_s$), we only find a weak $n-M_*$ correlation ($r_s = 0.27-0.30$)  for the red, green, and blue galaxies at $0.5 < z < 1$.  The blue and green galaxies are found to have similar S\'{e}rsic indices ($n \sim 1.5$) at higher redshifts ($z>2$). 
However, for galaxies at $1.5<z<2.5$, the shape of their profile seems to be insensitive to the stellar mass, with $|r_s|<0.05$. 


The green galaxies are found to have intermediate S\'{e}rsic indices at $0.5 < z < 2$.
It suggests that quenching process is accompanied by the transformation of galaxy structural appearances for galaxies at $z<2$ within a wide mass range. 
As galaxies move from BC to RS, their overall light profiles become cuspier (i.e., with a higher $n$). Previous works point out that quenching mechanisms are tightly coupled to the bulge (e.g., \citealt{Cameron+09, Wuyts+11a, Bluck+14}). 
Assuming the S\'{e}sic index $n$ is a proxy for bulge-dominance, it is found that galaxy quenching is accompanied with the buildup of bulge component.  


\subsection{The Mass Dependence of Galaxy Compactness}
The stellar mass density of galaxies has been supposed to connect the quiescent levels, and the surface density shows a strong correlation with the stellar population (\citealt{Bell+12, Lang+14, Whitaker+17}). 
\cite{Omand+14} show that quenching depends both on the stellar mass and the effective radius in the local Universe. Particularly, the quiescent fraction depends on the quantity $M_*r_{\rm e}^{-\alpha}$. 
\cite{Newman+12} derive $\alpha ^{-1}= 0.59-0.69$ for the QGs at $0.4 < z < 2.5$. Based on the observed trend in the size-mass relation for QGs, \cite{Barro+13} define the global degree of galaxy compactness (hereafter $\Sigma_{1.5}$) as the stellar mass divided by the effective radius to the power of 1.5: 
\begin{eqnarray}
\Sigma_{1.5} \equiv M_*r_{\rm e}^{-1.5}[M_\sun~{\rm kpc}^{-1.5}].
\end{eqnarray} 
This quantity lies between the surface stellar density, $\Sigma_e \propto M_*r_{\rm e}^{-2}$, and the inferred velocity dispersion, $\sigma^2 \propto M_*r_{\rm e}^{-1}$. These two quantities (i.e., $M_*r_{\rm e}^{-1}$ and $M_*r_{\rm e}^{-2}$) are confirmed to  be strongly correlated with color and SFR up to high redshifts (e.g., \citealt{Franx+08, Patel+13, Omand+14, Barro+17}). 
However, it is found that $\Sigma_{1.5}$ is the best one-dimensional description of quiescent fraction \citep{Omand+14}.  

\begin{figure*}
\includegraphics[scale=0.56]{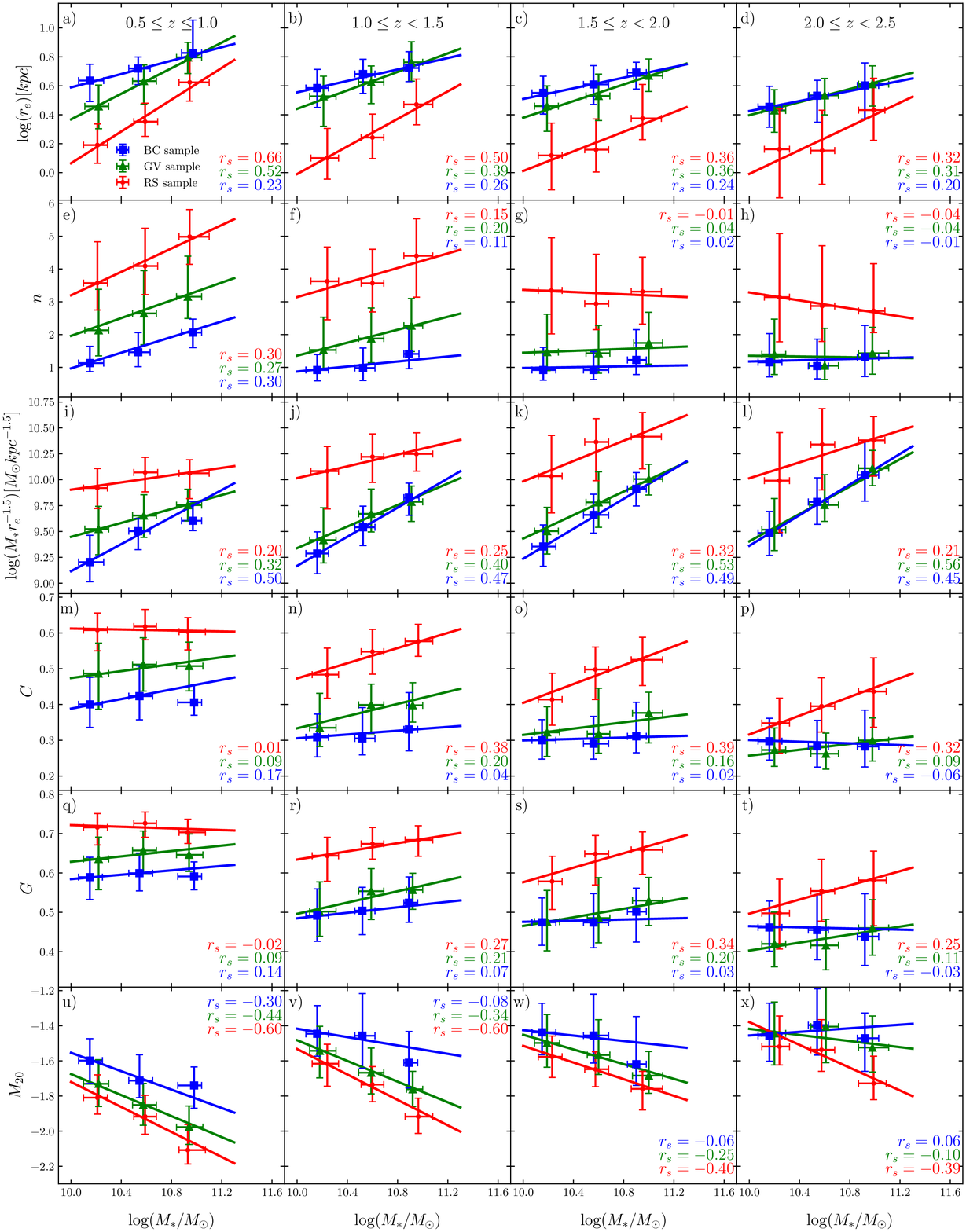}
\caption{From top to bottom: the structural parameters, $r_{\rm e}$, $n$, $\Sigma_{1.5}$, $C$, $G$, and $M_{20}$, as a function of the stellar mass for the RS (red), GV (green), and BC (blue) galaxies at $0.5 < z < 2.5$. For each galaxy population, four redshift bins and three stellar mass bins are considered. In each redshift and mass bin, the data point and error bar (RS: red dot; GV: green triangle; BC: blue square) represent the median and the 25th to 75th percentiles of the distribution for each sub-sample, respectively.
For each galaxy population in a redshift bin, the Spearman's rank correlation coefficient ($r_s$) is given together with a linear fit (solid line).
\label{fig04}}
\end{figure*}

The third line of Figure \ref{fig04} shows the $\Sigma_{1.5}$ distribution as a function of stellar mass for the three galaxy populations in four redshift bins. 
The blue galaxies at $0.5<z<2.5$ show very clear correlations ($r_s > 0.4$) between $\Sigma_{1.5}$ and stellar mass, which can be expected by the size-mass relations  shown in first line.  The weak mass dependence of size for blue galaxies means that mass assembly in blue galaxies could not effectively cause the growth of radius on average, which will lead to an increase of $\Sigma_{1.5}$.
The green galaxies also show a moderate correlation except at $0.5<z<1$. 
Compared with blue galaxies, the green galaxies at higher redshifts seem to have more significant and similar mass dependence of galaxy compactness. 
The red galaxies in four redshift bins are found to have the weakest $\Sigma_{1.5} - M_*$ correlation. 

For a specified mass bin, the typical $\Sigma_{1.5}$ value of red galaxies is significantly higher than those of blue and green galaxies.  
It suggests that galaxies have a large global compactness when they are completely quenched, which supports that $\Sigma_{1.5}$ is a good indicator of quiescence. 
It has been supposed that quenching mechanisms such as AGN and stellar feedbacks become more efficient when the stellar density reaches a certain threshold \citep{vD+15, Whitaker+17}.
Although there is not an explicit threshold of quiescence, what would be expected is that galaxies with higher $\Sigma_{1.5}$ levels are more likely to be quenched.

The $\Sigma_{1.5}$ difference between blue and red galaxies is more prominent at lower end of mass. 
For the lower mass bin, the green galaxies have the higher $\Sigma_{1.5}$ than the blue galaxies at $0.5 < z < 2$. The green and blue galaxies tend to have similar $\Sigma_{1.5}$ distributions at higher redshifts and at high mass end. 

\subsection{The Mass Dependence of Nonparametric Measurement}
Nonparametric measurements are advantageous for galaxies at high redshifts, where a larger fraction of them may be irregular and do not have distinct centers (see the review by \citealt{Conselice+14}). 
The last three lines of Figure \ref{fig04} show the distributions of these nonparametric measurements ($C$, $G$, and $M_{20}$) as functions of stellar mass for the massive galaxies in four redshift bins. 

In general, these is a similar behaviour between the concentration index ($C$) and Gini coefficient ($G$) over all redshift ranges.
No significant $C-M_*$ and $G-M_*$ correlations are found for the green and blue galaxies over all redshift ranges, with absolute value of correlation coefficients less than 0.3. 
The quiescent galaxies with prominent bulge components (i.e., S\'{e}rsic index $n>2.5$) are commonly found to have larger values of $C$ and $G$. 
Just as \cite{Peth+16}  pointed out, for galaxies with $n<3$, their concentration indices and Gini coefficients are proved to be sensitive to the S\'{e}rsic index $n$. 
The values of $C$ and $G$  tend to be a large constant (i.e., $C \sim 0.65$ and $G \sim 0.72$) for the bulge-dominated galaxies with $n>3$. 
For the red galaxies at $1<z<2$, their  $C$ and $G$ values are weakly correlated with stellar mass ($|r_s| \sim 0.3-0.4$). The red galaxies at $z<1.0$ have similarly large values of $C$ and $G$, with no $C-M_*$ and $G-M_*$ correlations. 

As to the $M_{20}$ distributions, more massive galaxies tend to have lower $M_{20}$ values on average. 
In general, the $M_{20}$ difference between blue and red galaxies is more significant at higher redshift. 
The red galaxies have lower median of $M_{20}$ than the blue galaxies over whole ranges of redshift and mass, which points to an overall trend that baryonic matters in blue galaxies are gradually migrating to the center.  
For the red, green, and blue galaxies at $0.5 < z < 1$,  correlations between $M_{20}$ and stellar mass can be clearly found with  $|r_s| > 0.3$. 
The $M_{20}$ values of red galaxies exhibit anti-correlations with stellar mass, with the largest absolute values of correlation coefficient ($|r_s| = 0.35 - 0.57$).  The green galaxies show a weaker correlation between $M_{20}$ and mass. For the blue galaxies at $z>1$, no mass dependence of  $M_{20}$ can be found.

\section{Relation between SFR and stellar mass}\label{sec:SFR-M-trans}
The relation between star formation history and morphology since $z \sim 2.5$  has been studied (e.g., \citealt{Wuyts+11b, Lee+18}). 
In this section, we explore how the colors and morphologies of massive galaxies correlate with their locations in the SFR$-M_*$ diagram.
\subsection{SFR$-M_*$ Relations for BC, GV, and RS Populations }
\begin{figure*}
\includegraphics[scale=0.60]{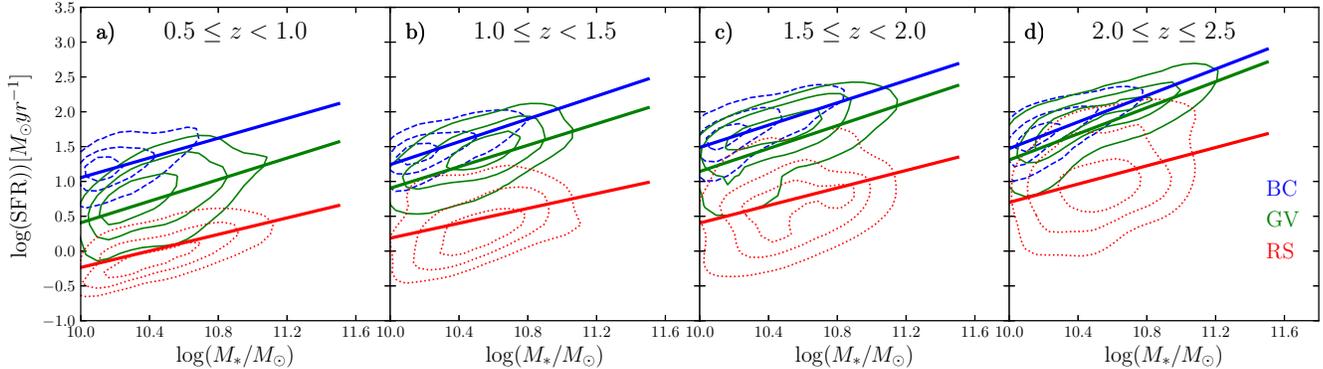}
\caption{Distributions for the three galaxy populations in SFR-$M_*$ plane in four redshift bins. The red dotted, green solid, and blue dashed contours represent the red, green, and blue galaxy populations, respectively, and contours contain 20\%, 50\%, and 80\% of data points from inside to outside. The linear fits for the three galaxy populations are also shown as the heavy solid lines.
\label{fig05}}
\end{figure*}

At first we present the contour maps in the SFR-$M_*$ diagram for the three galaxy populations in four redshift regions in Figure \ref{fig05}. 
The linear fittings for BC, GV, and RS galaxies are also given. The best-fitting coefficients for the three galaxy populations in four redshift slices are tabulated in Table \ref{tab:coeff}. 
Generally, a rising trend of the overall SFR along redshift is very clear (e.g, \citealt{Speagle+14, Whitaker+14}). 
The slope tends to be larger at higher redshifts for all populations. 
For the blue galaxy population, the intercepts do not change too much, which is consistent with the peak of star-forming rate density at $z\sim2$ (\citealt{Madau+14}). 
It is found that the GV population is located in an intermediate region, between BC and RS, on the SFR-$M_*$ diagram. 
The sequence moving from BC to RS is a sequence that SFR decreases. Although both SFR and dust corrected U-V color are correlated with the amount of recently formed stars relative to the evolved, we stress that they will have the different changes as the amount of young population increases. 
At $z \sim 2$, it is found that the GV contours tend to overlap with the BC contours, which implies that it is hard to distinguish the green galaxies from blue galaxies at early epoch. The reason might be that the bimodality is weaken at higher redshifts.  Another alternative is also likely due to the greater uncertainty in determinations of SFR and dust attenuation at higher redshifts. 

\begin{deluxetable*}{cccccccccc}
\tablecaption{The coefficients of linear fits for BC, GV, and RS populations in four redshift slices, where the error bars are 95\% confidence limits. 
\label{tab:coeff}}
\tablehead{\colhead{redshift range} & \multicolumn2c{Blue galaxies} & \multicolumn2c{Green galaxies} & \multicolumn2c{Red galaxies} \\
\colhead{} & \colhead{$\alpha \pm \Delta \alpha$} & \colhead{$\beta \pm \Delta \beta$ } & \colhead{$\alpha \pm \Delta \alpha$} & \colhead{$\beta \pm \Delta \beta$ } &  \colhead{$\alpha \pm \Delta \alpha$} & \colhead{$\beta \pm \Delta \beta$}
}
\startdata
$0.5\leqslant z < 1.0$         &$0.71\pm0.10$&$1.05\pm0.04$& $0.77\pm0.11$ & $0.41\pm0.06$ & $0.59\pm0.07$  & $-0.24\pm0.04$ \\
$1.0\leqslant z < 1.5$         &$0.82\pm0.07$&$1.24\pm0.03$& $0.77\pm0.11$ & $0.90\pm0.06$ & $0.53\pm0.10$  & $0.19\pm0.06$ \\
$1.5\leqslant z < 2.0$         &$0.81\pm0.07$&$1.48\pm0.03$& $0.83\pm0.11$ & $1.13\pm0.07$ & $0.63\pm0.12$  & $0.40\pm0.08$ \\
$2.0\leqslant z \leqslant 2.5$ &$0.95\pm0.07$&$1.47\pm0.03$& $0.98\pm0.11$ & $1.27\pm0.07$ & $0.69\pm0.19$  & $0.67\pm0.12$ \\
\enddata
\tablecomments{The SFR$-M_*$ relations for three populations  are parameterized as $\log({\rm SFR}) = \alpha [\log (M_*/M_\sun) - 10.0] + \beta$.
}
\end{deluxetable*}

\subsection{SFR$-M_*$ Relations for Different Morphologies}
It has been found that these is a strong correlation of galaxy structural properties with their relative deviations from the SFMS on the SFR-$M_*$ plane (\citealt{Wuyts+11b, Brennan+17,Lee+18}). A nearly monotonic trend towards higher $n$, smaller $r_{\rm e}$, and higher stellar density is found as galaxies move away from the SFMS.  The semi-analytic model (SAM) can reproduce this trend, in which the bulge growth driven by mergers and disc instabilities is accompanied by the growth of central supermassive black hole, and the AGN feedback may regulate or quench star formation activities (\citealt{Brennan+17}). 
We already find that the blue galaxy population is dominated by late-type disk (LTD) and irregular (IRR) galaxies, whereas the red galaxy population is dominated by spheroid/bulge dominated (SPH) galaxies (\citealt{Gu+18}). 
The correlation between morphology and galaxy population is  remarkable, which is consistent with the morphology -- sSFR relation in the local universe \citep{Bait+17}.
The green galaxies are often considered to be under transition from star forming phase to quiescent one. 

\begin{figure*}
\includegraphics[scale=0.7]{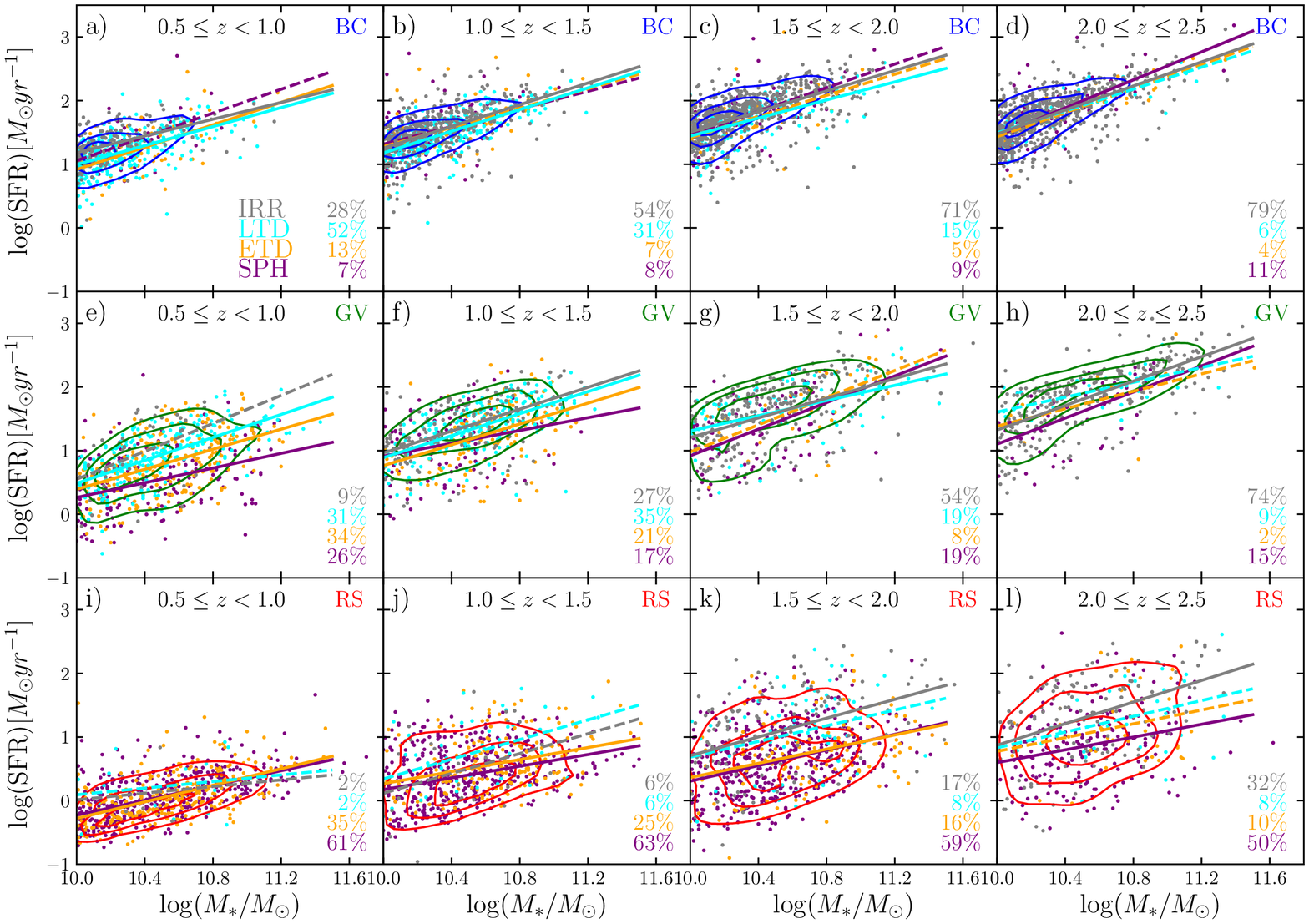}
\caption{Morphological distributions for three galaxy populations on SFR-$M_*$ plane in four redshift bins. Locations and proportions of galaxies with the SPH, ETD, LTD, and IRR morphologies are represented in pink, orange, cyan, and gray colors, respectively. We overlap the same contours of BC, GV, and RS galaxies as in Figure \ref{fig05}. The  linear fittings for the morphological subsamples are also given. Dashed lines represents the subsamples containing a small fraction ($\leqslant 10\%$) of galaxies, and solid lines for the big subsamples. 
\label{fig06}}
\end{figure*}

The morphologies for  green galaxies are various, and are intermediate between those of RS and BC, which is consistent with some previous works (\citealt{Mendez+11, Salim+14, Ichikawa+17}), suggesting that  quenching processes are accompanied by the buildup of bulge. 
With a large sample of green galaxies at $0.5<z<2.5$ in 3D-$HST$/CANDELS, we can observe how their morphologies transform along a sequence from the late- to early-type galaxies (i.e., from IRR to LTD to ETD to SPH) when they move far away from the SFMS. 
The galaxy morphologies and locations of the three galaxy populations on the SFR$-M_*$ plane are shown in Figure  \ref{fig06}.  The galaxies with different morphologies are denoted in different colors. 
Our linear fitting results for the green galaxies with different morphologies are also shown in each redshift slice. 

The morphological transformations from IRR to SPH may reflect  the process of bulge buildup during the star-formation quenching. Such a morphological change can be explained by some violent and rapid processes, such as galaxy mergers (\citealt{T&T+72}) or protogalactic collapse due to disc instability (\citealt{Dekel+09}). 
An alternative interpretation of morphological transformation is  internal and environmental secular processes, such as  bar instabilities, galactic wind, minor mergers, and harassment (\citealt{KK+04}). 
It can be seen in Figure \ref{fig06} that a vast majority of BC population is late-type galaxies whereas a majority of RS population are early-types. 
In particular, the blue galaxies at $z>1$ are predominated by the irregulars. The blue galaxies with various morphologies have very tight SFR$-M_*$ correlations.
At $0.5<z<1.0$, the green galaxies with irregular morphologies have higher SFRs, which is close to the SFMS of blue SFGs (see Figure  \ref{fig06}).

For GV population, the SPH galaxies are found to have a lower mean SFR in the SFR distributions. 
We find a weak, but clear trend at $0.5<z<1.5$ that the SFR$-M_*$ relations evolve systematically along the morphological sequence, with a decreasing SFR trend  from IRR to SPH. It suggests a gradual transformation of morphology at low redshifts. 
For the green galaxies at higher redshifts ($z>1.5$), it is hard to distinguish the difference between the SFR$-M_*$ relations for the SPH, ETD, LTD, and IRR galaxies. 
This suggests a very short time scale of morphology transformation at higher redshifts. 
There is a clear trend that the IRR fraction for GV population increases dramatically with redshift. 
At $1.5<z<2.5$, more than 50\% of green galaxies are found to be irregular, indicating a higher probability of major merger or disc instability during star formation quenching at $z \sim 2$. 
Such violent events may drive AGN and/or supernova feedbacks, and play an important role in the rapid quenching of star formation. 
At lower redshifts ($0.5<z<1.5$), more scattered SFR distributions can be found for green galaxies, which perhaps implies a longer time scale of morphology transformation at lower redshifts. 


\section{Discussion}\label{sec:discuss}

\subsection{Structure Transformation and Quenching} \label{sec:mass-morph}
\begin{figure*}
\includegraphics[scale=0.45]{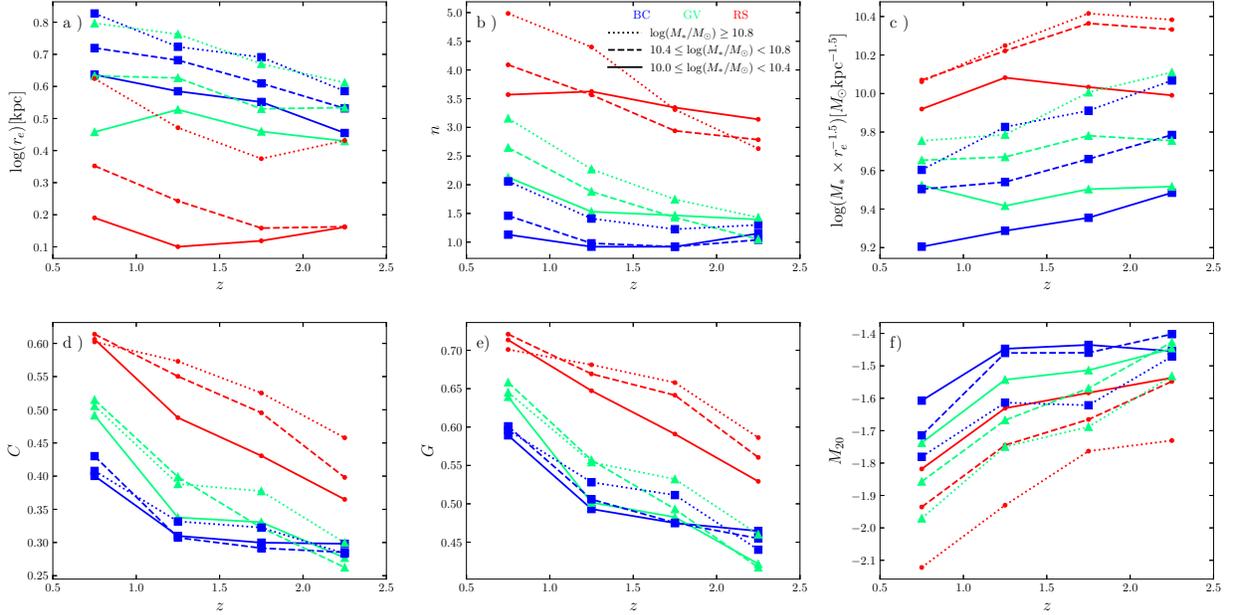}
\caption{Median values of the structural parameters ($r_{\rm e}$, $n$, $\Sigma_{1.5}$, $C$, $G$, and $M_{20}$) as a function of stellar mass and redshift for blue galaxies (blue squares), green galaxies (green triangles), and red galaxies (red dots). The stellar mass bins are $10.0 \leqslant \log(M_*/M_\sun) < 10.4$ (solid lines), $10.4 \leqslant \log(M_*/M_\sun)< 10.8$ (dashed lines), and $\log(M_*/M_\sun) \geqslant 10.8$ (dotted lines).
\label{fig07}}
\end{figure*}
Histories of stellar mass assembly for galaxies are mainly classified into two distinct modes: ``inside-out'' and ``outside-in'' \citep{Pan+16, WangEC+17}. In the ``inside-out'' assembly mode, star formation quenching takes place first in the center region of a galaxy,  which may be due to AGN feedback depleting, heating or blowing out the gas in galactic center (e.g., \citealt{Kauffmann+04}, \citealt{Croton+06}, \citealt{Fabian+12}, \citealt{Dekel+14}) or the bulge stabilizing gas in the disk from collapsing (i.e., so-called ``morphological quenching''; \citealt{Martig+09}).  In the outside-in assembly mode, the cessation of star formation starts on the outskirt region, which may be due to the environmental effects striping the gas from the outskirts , such as gas stripping and galaxy harassment \citep{Moore+96, GG+72, Guo+17, Papovich+18}. 
These two assembly modes are expected to change the light profile of a galaxy in a different way during its quenching process (from BC to GV to RS). 
The evolution of massive galaxies can be regarded as a process of mass assembly. 
By assuming that green galaxies are in the transitional phase from star-forming to quiescent status,  the changes of structural parameters for blue, green, and red galaxies may reflect the mass dependence of morphological transformations during star formation quenching.  

Observations reveal the morphology/structure -- SFR/color relations (especially at a fixed stellar mass) (e.g., \citealt{Cameron+09}, \citealt{Wake+12}, \citealt{Omand+14}, \citealt{Bluck+14, Bluck+16}, \citealt{Woo+15}, \citealt{Teimoorinia+16}). 
In general, we also find that there is a clear offset in structure and star formation rate between massive blue and red galaxies. The morphological distinctions between blue and red galaxies are a manifestation of structural transformation during the quenching phase. 
Compared with the blue galaxies, the red galaxies at $0.5 < z < 2.5$ in our sample are found to have smaller effective radius ($r_{\rm e}$), larger S\'{e}rsic index ($n$), larger galaxy compactness ($\Sigma_{1.5}$), lower concentration ($C$), lower Gini coefficient ($G$), and lower second-order moment ($M_{20}$). 
For the green galaxies at $0.5 < z < 2.0$, the overall distributions of structural parameters in each mass bin are intermediate between those of red and blue galaxies. This points to that morphological transformation (e.g., bulge buildup) in the transitional phase is accompanied by the quenching process, which is consistent with our previous work (\citealt{Gu+18}).
For the massive ($\log M_*/M_{\sun} > 10.8$) blue and green galaxies, their sizes and compactness tend to be similar, and they show quite distinct structure distributions from red galaxies. 
It suggests that the inside-out stellar mass assembly mode dominates in massive SFGs at higher redshifts, and the change in effective radius will not be remarkable when the SFGs are partly quenched from inside/central regions. 
These massive galaxies will not appear to be significantly shrunk until they are completely quenched into red QGs.

\subsection{Redshift Evolution of Structural Parameters}
Figure \ref{fig07} shows the redshift evolutions of the median values of structural parameters for the blue, green, and red galaxies within three mass bins: $10.0 \leqslant \log(M_*/M_\sun) < 10.4$ (low mass bin), $10.4 \leqslant \log(M_*/M_\sun)< 10.8$ (middle mass bin), and $\log(M_*/M_\sun) \geqslant 10.8$ (high mass bin). 
As expected, the coverages of $r_{\rm e}$ and $M_{20}$  are wider for the red galaxies at lower redshifts, and compactness $\Sigma_{1.5}$ for blue galaxies have a considerable variation from low- to high-mass ends. 
Interestingly, over the whole redshift range ($0.5<z<2.5$), the blue and red galaxies in low mass bin are found to have constant S\'{e}sic indices:  $n \sim 1$ for blue galaxies, and $n \sim 3.5$ for red galaxies. 
The red galaxies in low mass bin show remarkable redshift evolutions of the nonparametric measurements $C$ and $G$.

The panel (c) of Figure~\ref{fig07} shows a rising trend of compactness $\Sigma_{1.5}$  with redshift. The blue SFGs at low redshifts have the lowest $\Sigma_{1.5}$, whereas high-$z$ red galaxies tend to be compact. This trend may help us to understand the rarity of compact SFGs at lower redshifts and the prevalence of compact QGs at $z \sim 2$ \citep{Lu+19}.  
The panel (f) shows that the high-$z$ SFGs with lower stellar masses are common to have the highest $M_{20}$, which corresponds to very extended structures.

In Figure \ref{fig07}, it is clear to show that green galaxies are intermediate in structure between red and blue galaxies in each stellar mass bin at $z < 2$, which supports that green galaxies are in a transitional phase.
Although similar $r_e$ and $\Sigma_{1.5}$ for BC and GV galaxies is found at high mass end, it suggests the inside-out stellar mass assembly mode dominates in massive galaxies, which also discuss in Section \ref{sec:mass-morph}. 

\subsection{Transition Time-Scale as a Function of Redshift}\label{sec:timescales}
To back up our opinion of morphological transformations, we estimate the average transition time-scale as a function of redshift,  following \cite{Pandya+17}. 
Above all, it needs to make a strong assumption that the quenching process is a monodirectional track from star-forming galaxies to quiescent galaxies. In this paper, we regard the green and red galaxy populations as the transition and quiescent populations, respectively. Based on the cubic polynomial fits to the observed number densities of our BC, GV, and RS galaxies, the average transition time-scale can be estimated as the following definition: 
\begin{equation}
 \langle t_{\rm transition} \rangle _{z_1,z_2} = \langle n_{\rm transition} \rangle _{z_1,z_2}{\left(\frac{{\rm d}\,\,n_{\rm quiescent}}{{\rm d}\,t}\right)}^{-1}_{z_1,z_2},
\label{eq:timescale}
\end{equation}
where $\langle t_{\rm transition} \rangle _{z_1,z_2}$ and $\langle n_{\rm transition} \rangle _{z_1,z_2}$ are the average transition time-scale and the average number densities of green galaxies between two closely spaced redshifts, $z_1$ and $z_2$, 
and ${\left(\frac{{\rm d}\,\,n_{\rm quiescent}}{{\rm d}\,t}\right)}^{-1}_{z_1,z_2}$ is a variation in number density of red galaxies over a period of cosmic time between $z_1$ and $z_2$. 
It is noticed that if the quenching process is not a simple monodirectional track, the galaxies can move across the middle region on SFR$-M_*$ plane more than one time (e.g., rejuvenation events or SFMS oscillations). The above measurement of transition time-scale is still meaningful, which can be interpreted as how long galaxies averagely stay in the transition region in total. If we remove the strong assumption of the monodirectional evolutionary track, the average transition time-scale defined here can be regarded as the upper limit on the average transition time-scale, as a function of redshift. 

\begin{figure}
\centering
\includegraphics[scale=0.90]{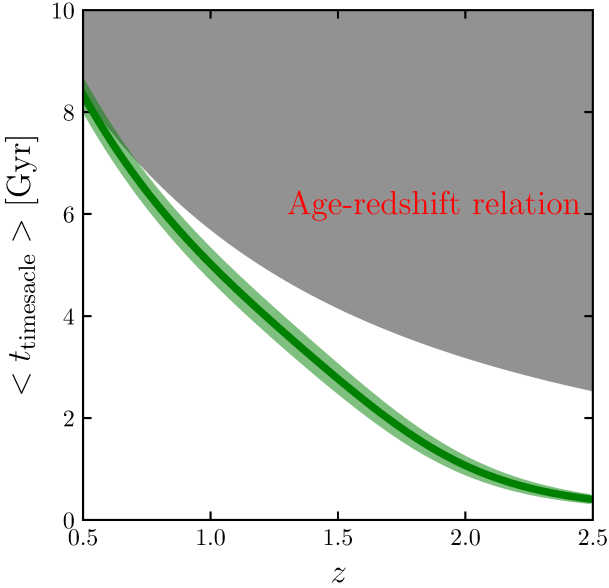}
\caption{Observational upper limit on the average population transition time-scale as a function of redshift (green solid line with shadow representing the propagated errors). The gray area is the area above the age-redshift relation. 
The average transition time-scale is increasing with cosmic time, 
which is consistent with the fast track dominated at high redshift, whereas the slow track dominated at low redshift. 
\label{fig08}}
\end{figure}

The number densities of galaxies in BC, GV, and RS populations as a function of redshift are shown in Figure  \ref{fig03}. 
We divide our sample into 12 redshift bins. 
The results of the cubic polynomial fits are used to estimate the average population transition time-scale  using Eq (\ref{eq:timescale}). 
For the massive galaxies  ($M_*> 10^{10} M_\sun$), the upper limit on the average transition time-scale as a function of redshift is shown in Figure  \ref{fig08}. The average transition time-scale increases smoothly with the cosmic time. As expected, galaxy quenching is on a fast track at high redshifts, whereas on a slow track at low redshifts. 
Given that the quenching processes are accompanied by the morphological transformation, the transition time-scale trend is consistent with the difference of the morphological transformation, as described in Section \ref{sec:SFR-M-trans}. 
It is a natural consequence that a fast morphological transformation corresponds to a short transition time-scale. 
It should be emphasized that our calculation is only for the massive galaxies with $M_*> 10^{10} M_\sun$. 

\subsection{The Mass-Matched Sample} \label{sec:mass-match}
Based on a large sample of red, green, and blue galaxies at $0.5<z<2.5$,  
the effective radius ($r_{\rm e}$), the compactness ($\Sigma_{1.5}$), and the second order moment ($M_{20}$) are found to be correlated with the stellar mass.
More specifically, at least one of the three galaxy populations in a specified redshift region shows a clear mass dependence ($|r_s| > 0.4$) of the above-mentioned three parameters:  $r_{\rm e}$ (for the red galaxies at $0.5<z<1.5$ and the green galaxies at $0.5<z<1.0$), $\Sigma_{1.5}$ (for all the blue galaxies and the green galaxies at $1.5<z<2.5$), and $M_{20}$ (for the red galaxies at $0.5<z<1.5$ and the green galaxies at $0.5<z<1$). 
On the other hand, the other parameters ($n$, $C$, and $G$) show no or very weak correlations ($|r_s| < 0.4$) with stellar mass over the whole redshift range. 
Therefore, we should take special care of stellar mass distribution when discussing cosmic evolution of the following three structural parameters: $r_{\rm e}$, $\Sigma_{1.5}$, and $M_{20}$.

\begin{figure*}
\centering
\includegraphics[scale=0.45]{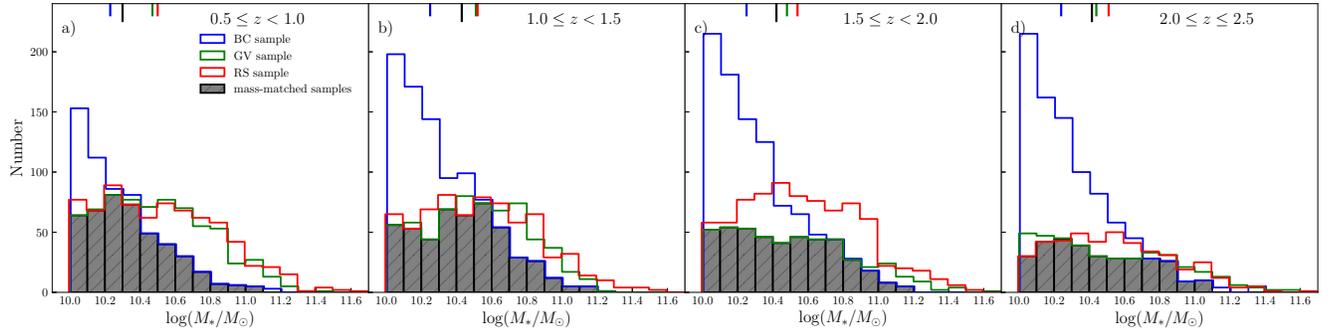}
\caption{Distributions of stellar mass in four redshift bins for the BC (blue), GV (green), RS (red) galaxies, and mass-matched samples (gray histograms), respectively. Color bars at the top of each panel represent the median values of stellar mass for the original and mass-matched samples. \label{fig09}
}
\end{figure*}

In order to minimize the stellar mass effects on the statistics of these three structural parameters (i.e., $r_{\rm e}$, $\Sigma_{1.5}$, and $M_{20}$), we build a ``mass-matched'' sample of massive galaxies at $0.5<z<2.5$ to ensure that the selected BC, GV, and RS galaxies have the same distributions of redshift and stellar mass. For a given redshift slice ($\Delta z = 0.1$) and mass interval ($\Delta \log(M_*/M_\sun)=0.1$) , the number of galaxies in the mass-matched sample  is defined by the minimum count among three populations. Specifically, for each galaxy with ($z_0$, $M_{*,0}$) in the galaxy population with the minimum count, the unique galaxy with the smallest value of $\sqrt{(z_i-z_0)^2+(\log M_{*,i}-\log M_{*,0})^2}$ is picked up from other two populations into the mass-matched sample. As a result, three populations in the mass-matched sample have the same galaxy count in each redshift bin, and have similar distribution of stellar mass.  Finally,  a mass-matched sample including 1686 galaxies for each galaxy population is achieved. 
Figure \ref{fig09} shows the distributions of stellar mass for the three galaxy populations in original (i.e., red, green, and blue lines) and mass-matched samples (i.e., gray histograms).  
The Kolmogorov-Smirnov (KS) tests are performed to check whether the $M_*$- and $z$-distributions of green galaxies in the mass-matched sample are similar to those of red and blue galaxies. Table \ref{tab1} lists the sample size and the KS probabilities.

\begin{figure*}
\includegraphics[scale=0.45]{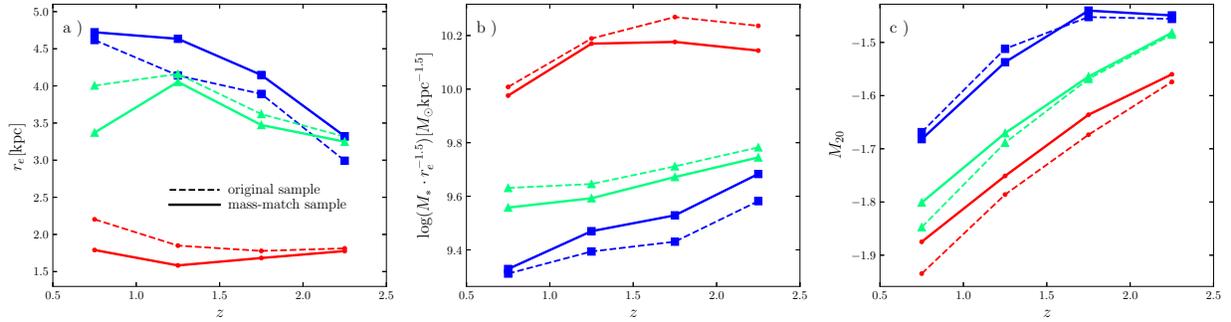}
\caption{Redshift evolutions of the median values of $r_{\rm e}$, $\Sigma_{1.5}$, and $M_{20}$. The thin dashed lines represent the results of the original samples and the thick lines represent the results of the mass-matched samples.
The blue square, green triangles, and red dots represent the BC, GV, and RS galaxies, respectively.    
\label{fig10}}
\end{figure*}

We compare the structural results of the mass-matched sample with those of the original samples. 
Figure \ref{fig10} shows the redshift evolutions of the median values of effective radius, compactness, and $M_{20}$.
The thin dashed lines represent results of the original samples, and the thick lines represent those of the mass-matched sample. 
The mass-matched sample gets rid of  excess proportion of low-mass blue galaxies, and the median value of stellar mass of blue galaxies is raised statistically. On the contrary, some high-mass green and blue galaxies are rejected, and their median values of stellar mass decrease. 
Thus, the median $r_{\rm e}$ of blue galaxies is significant larger than that of green galaxies at $z < 2$ in the mass-matched sample. However, it is still hard to differentiate the size distributions for blue and green galaxies at $z > 2$, which suggests that green and blue galaxies possess the similar sizes at $z > 2$.  
The compactness $\Sigma_{1.5}$ of red galaxies is still distinct from those of blue and red galaxies after mass matching. It should be noted that the median compactness for blue and green galaxies trend to converge at $z > 2$,  suggesting that green and blue galaxies may possess the similar structures at $z > 2$.  

As shown in Figure  \ref{fig04}, the values of $M_{20}$ are not strongly correlated with stellar mass for blue galaxies especially at high redshifts ($z>1.5$). 
Although the median $M_*$ value of blue galaxies increases after mass matching, the median $M_{20}$  value of blue galaxies, as a function of redshift, is still  not changed. 
For green and red galaxies, $M_{20}$ may anti-correlate with stellar mass. 
The green and red galaxies  in the mass-matched sample are found to have  systematically higher $M_{20}$ values. 


The three structural parameters (i.e., $r_{\rm e}$, $\Sigma_{1.5}$, and $M_{20}$) that correlate with stellar mass do have considerable changes.
It should take special care of stellar mass distribution when discussing and comparing the distributions of $r_{\rm e}$, $\Sigma_{1.5}$, and $M_{20}$. 
It is confirmed that green galaxies have intermediate structural parameters at $z < 2$ on the basis of the mass-matched samples, which supports that green galaxies are in a transitional phase. Additionally, 
blue and green galaxies are also confirmed to have similar structural parameters at $z>2$ by the statistics of the mass-matched sample. 

\begin{deluxetable}{cccccc}
\tablecaption{Sample size and results of the KS tests of the $M_*$- and  $z$-distributions  \label{tab1}}
\tablehead{
\colhead{redshift} & \colhead{galaxy} & \multicolumn2c{\underline{$z$-distribution}} & \multicolumn2c{\underline{$M_*$-distribution}} \\
\colhead{range} & \colhead{number} & \colhead{$P_{g-b}$} & \colhead{$P_{g-r}$} & \colhead{$P_{g-b}$} & \colhead{$P_{g-r}$} 
}
\startdata
0.5$-$1.0 & 440& 0.957& 0.688& 0.957 & 0.996 \\
1.0$-$1.5 & 491& 0.998& 0.891& 0.998 & 0.987 \\
1.5$-$2.0 & 438& 0.977& 0.999& 1.000 & 0.999 \\
2.0$-$2.5 & 317& 0.781& 0.692& 0.994 & 1.000 \\
\enddata
\end{deluxetable}

\section{Summary}\label{sec:summary}
By combining the multi-wavelength data in five 3D-$HST$/CANDELS fields, we construct a sample of massive ($M_* \geqslant 10^{10} M_\sun$) galaxies at $0.5 \leqslant z \leqslant 2.5$, covering $\sim 900\ \rm arcmin^2$ in total. 
Based on the extinction-corrected rest-frame $U-V$ color, we separate galaxies in our sample into the BC, GV, and RS galaxies up to $z \sim 2.5$ (the original sample, \citealt{Gu+18}).
The mass dependence of the structural parameters (i.e., $r_{\rm e}$, $n$, $\Sigma_{1.5}$, $C$, $G$, and $M_{20}$) are studied for the massive blue, green, and red galaxies at $0.5< z< 2.5$. The morphologies and relatively distributions of the three galaxy populations on the SFR-$M_*$ plane are investigated, especially for green galaxies. 
Finally, we build a mass-matched sample to re-examine the redshift evolutions of structural parameters.  

Our conclusions are summarized as follows:

1. Over ranges of stellar mass ($M_* > 10^{10} M_\sun$) and redshift ($0.5<z<2.5$), some structural parameters (say, $r_{\rm e}$, $\Sigma_{1.5}$, and $M_{20}$) are found to be correlated with stellar mass, where at least one of three populations shows a clear correlation with stellar mass. However, the other structural parameters (say, $n$, $C$, and $G$) are likely to be insensitive to stellar mass, showing no/weak correlation with stellar mass for all three populations. 

2. It is confirmed by the the SFR$-M_*$ diagram that our selection criteria of green galaxies based on the extinction-corrected rest ($U-V$) color works well in sample construction.
For the green galaxies at $0.5<z<1.5$, the morphological map in the SFR--$M_*$ diagram shows a sequence of morphology transformation (from IRR to SPH, i.e., bulge growth)  as they deviate from the SFMS.  It indicates that rapid and violent events such as mergers and disc instability come to be dominant mechanism of morphological transformation at  $z \sim 2$, while secular processes might be responsible for morphology change at $0.5<z<1.5$.

3. There is a clear offset in structure and star formation rate between massive blue and red galaxies at a fixed mass bin,  suggesting that the structural transformation is accompanied by the quenching phase. The overall trend is that massive galaxies are being more compact and bulge-dominated (nucleated) during star formation quenching. 

4. The distinctions in $r_{\rm e}$ and $\Sigma_{1.5}$ distributions between blue and green galaxies begin to break down at the higher ends of mass and redshift. 
The similar sizes and compactness for the high-$z$ massive blue and green galaxies suggests that these galaxies will not appear to be significantly shrunk until they are completely quenched into red QGs.

5. For the green galaxies at $0.5<z<1.5$, a morphological transformation sequence of bulge buildup can be seen as they are gradually shut down their star formation activities, while a faster morphological transformation is verified for the green galaxies at $1.5<z<2.5$. Our results imply that secular processes take the control of morphological transformations at $0.5<z<1.5$, while mergers and/or violent disc instabilities come to be dominant at $1.5<z<2.5$. 

6. After minimizing the mass effect by using mass-match samples, we confirm that the GV population is a transitional population when blue galaxies are being quenched into red galaxies, since the morphological and structural parameters of green galaxies are found to be intermediate between those of blue and red galaxies at $z < 2$.

\acknowledgments
We appreciate the referee for many constructive comments that helped us substantially improve our paper. 
This work is based on observations taken by the 3D--{\it HST} Treasury Program (GO 12177 and 12328) with the NASA/ESA {\it HST}, which is operated by the Association of Universities for Research in Astronomy, Inc., under NASA contract NAS5-26555.
This work is supported by the National Natural Science Foundation of China (Nos. 11873032, 11673004, 11433005, 11173016) and by the Research Fund for the Doctoral Program of Higher Education of China (No. 20133207110006). Z.Y.C. acknowledge support from NSFC-11873045.




\end{document}